\documentclass[journal=jacsat,manuscript=article,maxauthors=99,etalmode=truncate]{achemso}
\usepackage[version=3]{mhchem}

\usepackage{braket}
\usepackage{hyperref} 
\usepackage{xcolor}
\usepackage{textcomp}
\usepackage{gensymb}

\author{Arsineh Apelian}
\affiliation{Materials Department, University of California, Santa Barbara, CA 93106-9510, U.S.A.}
\author{Mariya Romanova}
\affiliation{Department of Chemistry and Biochemistry, University of California, Santa Barbara, CA 93106-9510, U.S.A.}
\author{Vojtech Vlcek}
\alsoaffiliation{Materials Department, University of California, Santa Barbara, CA 93106-9510, U.S.A.}
\affiliation{Department of Chemistry and Biochemistry, University of California, Santa Barbara, CA 93106-9510, U.S.A.}
\email{vlcek@ucsb.edu}

\title[Shallow NV centers]
  {Surface coupling of NV centers over nanoscale lengths}

\begin{document}

%TC:ignore
\begin{abstract} 
Shallow nitrogen-vacancy (NV$^{-}$) centers in diamond are among the most promising quantum sensors, offering high sensitivity and nanoscale spatial resolution. These systems are, however, prone to decoherence due to coupling with surface states. Here, we study sub-surface NV$^{-}$ centers embedded into large diamond slabs (8 nm) using various surface orientations (100 and 111) and terminations (hydrogen and nitrogen terminators) and compute the quasiparticle states of the defect. Our results show how dynamical charge fluctuations near the surface influence defect stability. We find that the (100) N-terminated surface introduces strong surface-state instabilities, while the (111) N-terminated surface provides a more favorable configuration. However, many-body calculations (within the $GW$ approximation) reveal that defects placed shallower than $\sim$~4 nm are prone to surface-induced ionization. These findings establish an accurate theoretical limit on the minimum depth required for stable NV$^{-}$ centers, guiding the design of NV$^{-}$ based quantum sensors.

\end{abstract}
%TC:endignore

\vspace{1em} 
\vspace{1em}

\begin{figure}
    \centering
    \includegraphics[width=6in]{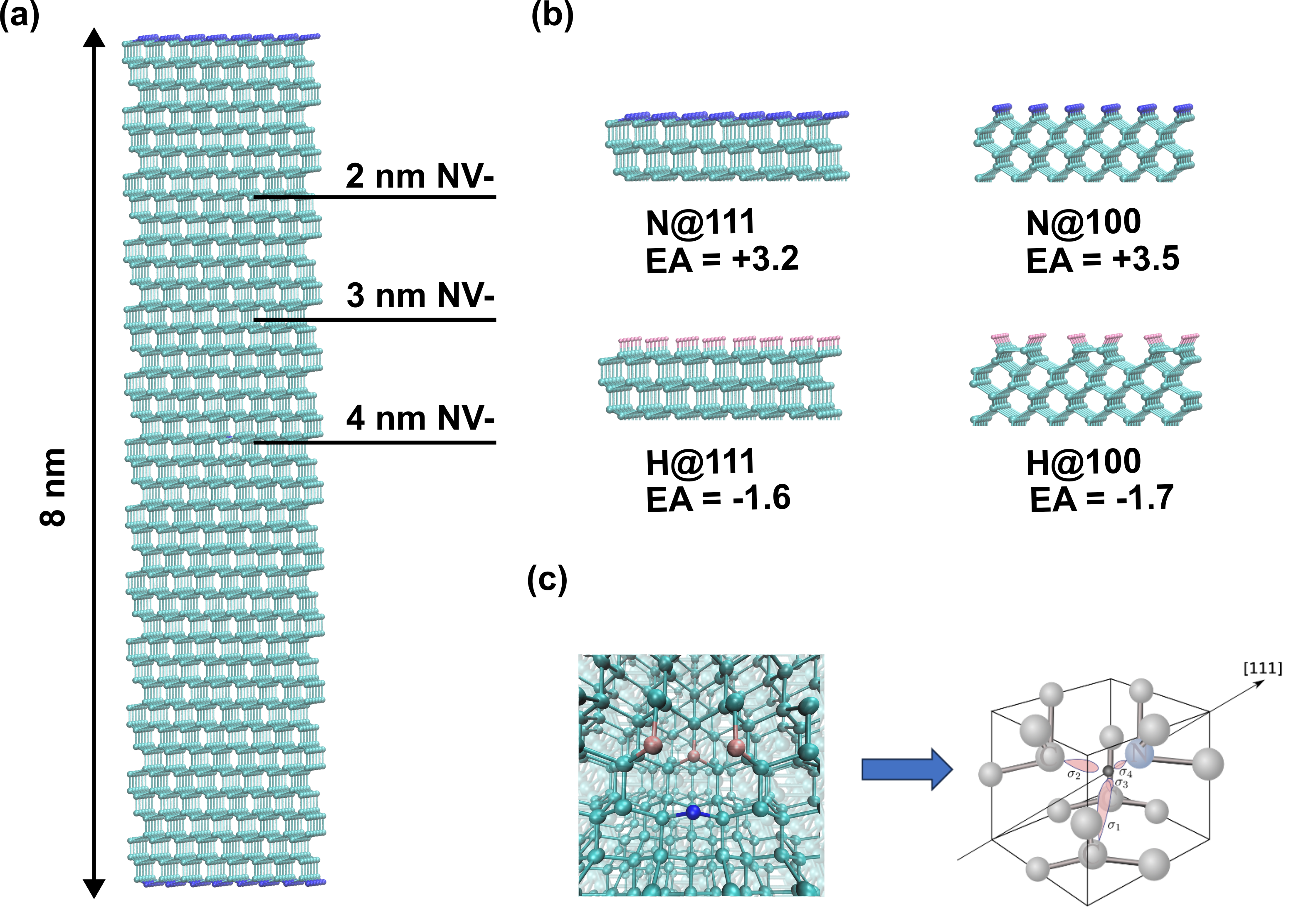}
    \caption{a) Example structure of an 8 nm slab nitrogen-terminated (111) surface with varying defect depths of 2 nm, 3 nm, 4 nm b) All four surface arrangements: (111) and (100) surfaces with nitrogen and hydrogen terminations with respective electron affinities taken from ref[\cite{AdamGali1}] and c) Schematic representation of the defect structure composed of carbon dangling bonds pointing towards the vacancy taken from ref[\cite{romanova}]} 
    \label{fig:1}

\end{figure}

The negatively charged nitrogen vacancy (NV$^{-}$) center in diamond hosts spin-locked states between valence and conduction bands, and can be implanted with sub-nm-scale accuracy in lateral dimensions and within nm accuracy in depth (i.e., below the surface) \cite{Awschalom2018, PhysRevB.74.161203, https://doi.org/10.1002/pssa.200671403}. With its paramagnetic ground state and spin-conserving optical transitions between ground and excited state triplets, the spins associated with the defect are sensitive to environment perturbations and external fields; consequently the NV$^{-}$ center is an excellent candidate for quantum sensing \cite{Balasubramanian2008, PhysRevLett.107.207210, Maze2008, Grinolds2013, PhysRevB.80.115202}. In fact, the NV$^{-}$ provides a unique combination of sensitivity and spatial resolution: a single NV$^{-}$ sensor can detect magnetic fields in the $m$T range with a $\mu$T uncertainty \cite{romanova, Balasubramanian2009} and electric fields in the $V/$cm range, both with nanometer-scale resolution and at room-temperature \cite{Dolde2011, Qiu2022}. The defect center has optical and magnetic properties that are determined by well-separated localized levels in the electronic band gap, which are dominated by localized orbitals pointing towards the vacancy center (Fig. \ref{fig:1}(c)). The precise energies of the single quasiparticle (QP) states in the gap of diamond determine the transition energies employed in sensing. Moreover, while there has been intensive research on NV$^{-}$ centers in bulk diamond, \cite{FJelezko_2004, https://doi.org/10.1002/pssa.200671403, PhysRevB.89.075203} no reliable results exists for accurate computation of excited-state transition energies and the role of surface coupling in shallow NV$^{-}$ centers. 

In practice, engineering the defect as close as possible to the surface i.e., a few nanometers ($<$10 nm) deep, ensures good sensing capabilities; however, this presents a significant challenge. Based on experimental observations \cite{PhysRevLett.113.027602, PhysRevLett.122.076101, PhysRevLett.115.087602, PhysRevLett.114.017601, PhysRevB.82.115449}, the energies of the electronic states associated with the defect are highly sensitive to charge fluctuations on the surface and, by extension, to the surface termination, local geometry, and defect depth---though the optimal configuration remains unknown. Hydrogenation and oxygenation are conventional treatments for diamond surfaces, but they affect near-surface NV$^{-}$ centers in distinct ways \cite{AdamGali1}. It has been widely reported that pure H-terminated surfaces lead to a negative electron affinity (NEA), inducing upward surface band bending and shifting the Fermi level below the acceptor level. \cite{PhysRevB.64.165411, 10.1063/1.3364135, PhysRevB.59.5847, petra, AdamGali1} Consequently, systems with a NEA can result in the desired negatively charged NV$^{-}$ center to be converted to a neutral NV$^{0}$ center. Additionally, the near-surface NV$^{-}$ centers may also suffer from surface state intrusion whereby an excited electron scatters to a surface-related image state, and gets emitted to the environment, resulting in blinking or permanent bleaching \cite{AdamGali1, doi:10.1021/nl501927y, PhysRevB.83.081304}.  On the contrary, calculations suggest \cite{AdamGali1,AdamGali2,PhysRevB.105.085305} that N-terminated surfaces have a large positive electron affinity (PEA) and, therefore, hybridizes less with the conduction states near the band gap, and hence have less influence on the in-gap defect states. Recent experimental works have achieved these types of surface alignments through different chemical vapor deposition techniques \cite{hughes2024stronglyinteractingtwodimensionaldipolar, ATTRASH2021121741, 10.1063/1.4930945, https://doi.org/10.1002/pssa.201532182}.

Understanding various couplings and control mechanisms is key to engineering improved quantum sensor networks; first-principles calculations can help identify and explain these interactions at the nanoscale. Calculations need to consider large slab models representing polarizable surface states that couple to the NV$^{-}$ \cite{item_1a009914bf47412fa998e7f270b9108a, PhysRevB.105.085305}. As a result, such investigations have been largely unexplored due to the exceedingly high computational cost associated with simulating thousands of atoms. Here, we overcome these computational bottlenecks and demonstrate realistic simulations of sub-surface NV$^{-}$ centers using mean-field and many-body perturbation theory (MBPT) in an 8 nm diamond slab containing approximately 15,000 electrons. The slab is large enough to capture the electronic structure of bulk diamond, and contains explicit surface terminations, e.g. hydrogen and nitrogen atoms, along the (111) and (100) directions with varying depths (e.g., 2 nm, 3 nm, 4 nm). We employ the stochastic formalism for the many-body theory to accurately predict individual electronic state energies that govern the NV$^{-}$ center’s functionality in quantum sensing applications.

\begin{figure}
    \centering
        \includegraphics[width=6in]{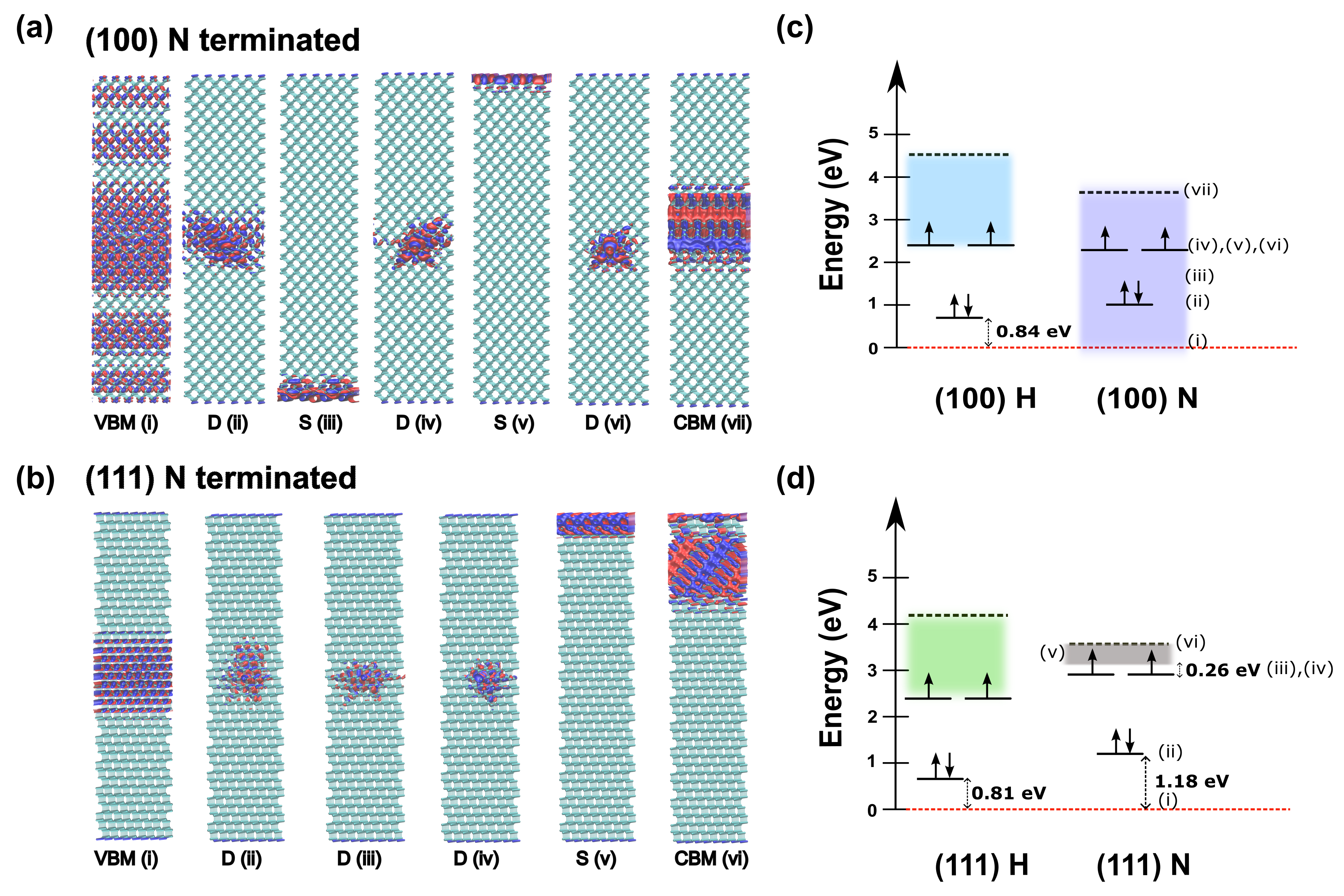}
    
    \caption{Relevant single-particle orbitals shown by isosurface plots: blue and red colors represent positive and negative values of real valued orbitals for the a) (100) N terminated slab and b) the (111) N terminated slab where we distinguish VBM and CBM states followed by the in-gap defect states and surface states. The surface states appear in between the defect states for the (100) N slab whereas the (111) N defect states are well isolated. (c) and (d) Computed electronic levels for all 4 systems. Energies are shifted by the VBM energy (shown as a red dashed line) and CBM is shown as a black dashed line. The colored blocks are the surface-related states: blue and green for (100) H and (111) H and purple and gray for (100) N and (111) N, respectively.} 
    \label{fig:2}

\end{figure}

We begin by inspecting the stability of the defect centers using Density Functional Theory (DFT). Since we are interested in quantitative predictions of QP levels (equivalent to the information obtained from photoemission spectra), the mean-field calculations serve as a starting point for the many-body calculations. We  investigate four systems in total: two surfaces with (111) and (100)  orientation, each with a nitrogen substitutional termination and a hydrogen passivation shown in Fig. \ref{fig:1}(b). For all systems, we begin by considering the NV$^{-}$ center placed at a 4 nm depth in an 8 nm length slab. Details on the construction of the slabs can be found in the SI. The DFT results are summarized in Fig.~\ref{fig:2} for all systems. 

For the (100) H-terminated system, our results show a significant number of surface states appear $\sim$~2.37 eV from the conduction band edge (indicated by the blue block in Fig.\ref{fig:2}(b)), but no surface states are present near the valence band edge. Similarly, for the (111) H-terminated system, our results reveal surface image states appear $\sim$~2.10 eV from the conduction band minimum (CBM) (represented as a green block in Fig. \ref{fig:2}(d)) and none from the valence band maximum (VBM). To our knowledge, all other studies of shallow NV$^{-}$'s observe surface states near the valence band, however, these calculations rely on diamond slabs that are much smaller with defects placed at 1 nm or less. We will revisit this point when we investigate the effects of positioning the defect at various depths. Still, we find that both (100) and (111) pure H-terminated diamond slabs are not ideal surface arrangements for quantum sensors. These observations are consistent with previous studies showing that H-terminated surfaces often suffer from charge state instability due to the system's NEA of -1.7 eV for (100) surface and -1.6 eV for (111) surface and appearance of deep surface-related states \cite{AdamGali1, AdamGali2, PhysRevB.83.081304, doi:10.1021/nl501927y}. The defect levels of the NV$^{-}$ can easily mix with these surface image states and any excitation of the NV$^{-}$ may lead to a temporary or permanent loss of the excited electron.

Previous calculations for the (100) N-terminated system indicate that it has a PEA of +3.5 eV, making it more likely for electrons to remain bound and maintain its negative charge state \cite{AdamGali2}. However, we find that in this system, the defect states are also heavily mixed with the surface states due to their near degeneracy with surface-related bands (shown as a purple block in Fig. \ref{fig:2}(b) which appear near the valence band by $\sim$~1.80 eV and the conduction band by $\sim$~1.55 eV. Surface states also appear in between the defect states, introducing midgap states and significantly interfering with the NV$^{-}$ center (see Fig. \ref{fig:2}(a)). This shows that while the N (100) diamond surface has a high PEA, surface states are still heavily influencing the instability of this system, making it the most unstable surface out of all four systems and, therefore, arguably the most unsuitable for sensing purposes.

Lastly, our results show that the (111) N-terminated slab is the most stable configuration. This is  consistent with the DFT calculations of Gali et al. \cite{AdamGali1,AdamGali2}, focusing on surface terminations without defects or, when including defects, used slabs with a thickness of $\sim$~2 nm. Other works \cite{PhysRevB.105.085305} also used mean-field methods and find the N-terminated (111) surface to be the optimal choice for quantum sensors, however, these calculations are for very shallow NV$^{-}$ centers (1.5 nm below the surface) due to computational costs. In our work, we find that only a few surface related bands appear $\sim$~0.45 eV from the CBM (shown as a gray block in Fig. \ref{fig:2}(d) and none near the VBM. The defect states are also well separated from the band edges compared to the other three systems: 1.18 eV from the VBM and 0.26 eV from the defect states to the first unoccupied surface state. All other systems have defect states that are nearly degenerate with the unoccupied surface states: the (100) N-terminated has the smallest separation with an energy difference of $<$ 0.1 meV  with strong surface mixing within the defect levels while the H-terminated surfaces are slightly better ($\sim$~0.4 meV for (100) and $\sim$~3 meV for (111)). The energy separation between the VBM and first defect state also decreases significantly comparing the (111) N with the other surface arrangements: 0.84 eV for (100) H-terminated, 0.81 eV for (111) H-terminated, and 0.03 eV for (100) N-terminated. Therefore, the (111) N-terminated diamond surface appears as ideal for improved quantum sensors.

\begin{figure}
    \centering
    \includegraphics[width=4in]{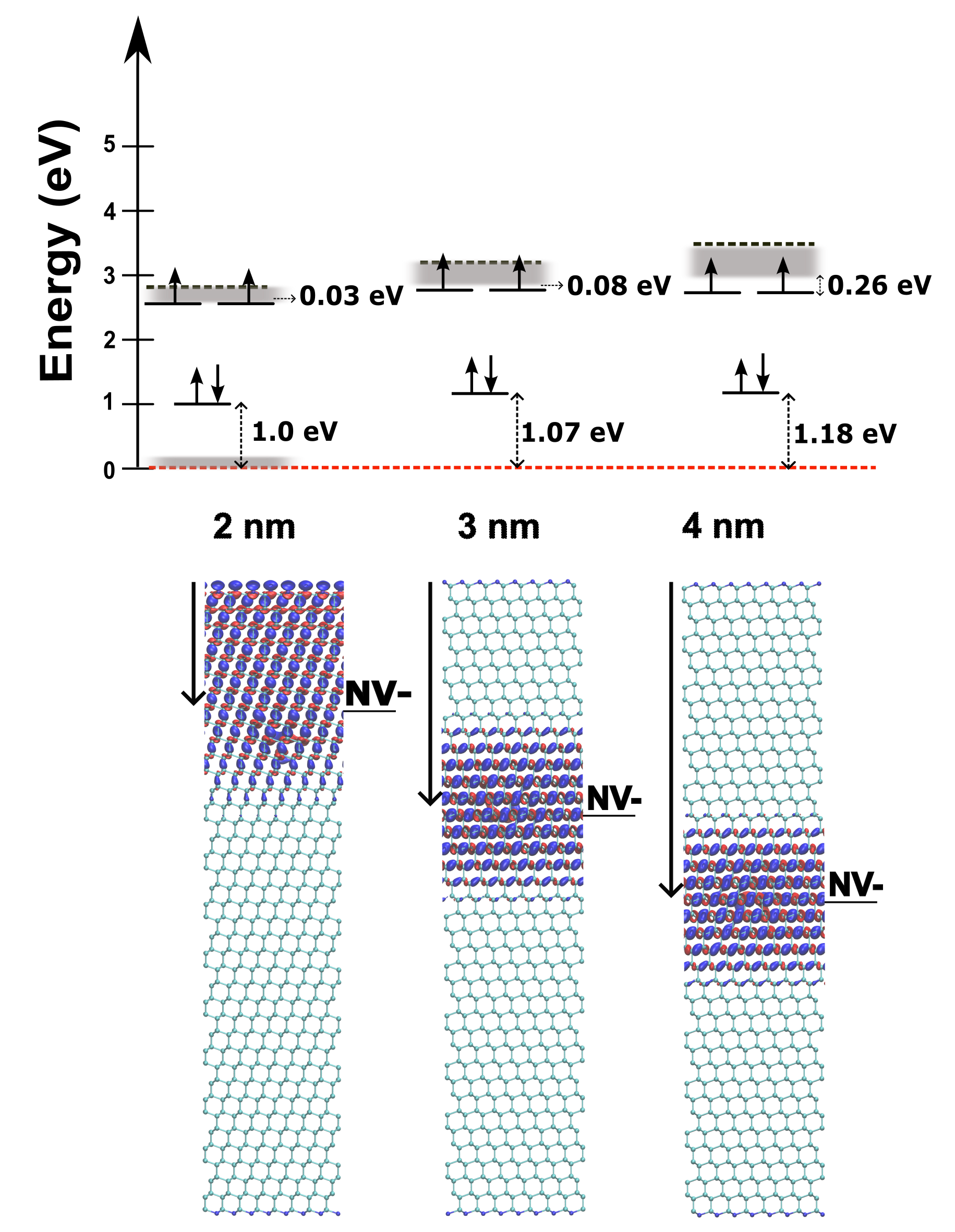}
    \caption{Energy diagram for the (111) N terminated slab with the defect center placed at: 2 nm, 3 nm, and 4 nm with respective plotted orbitals of the VBM state for all systems. Energy separations are shown between the VBM (red dashed line) and the first defect state as well as the two unpaired, degenerate defect states with the first unoccupied surface state. Surfaced bands are shown as gray blocks. The energy separation between the two unpaired defect states and the first unoccupied surface state gets reduced as the depth decreases. Additionally, a surface-bulk hybridized state appears for the 2 nm defect depth.} 
    \label{fig:3}

\end{figure}

Next, we investigate how defect center depths change the electronic levels (and surface states) and the overall stability of this system. 
Note that the NV$^{-}$ center is placed as close as possible to the surface to enhance the sensor’s sensitivity and spatial resolution. However, the precise distance at which the defect can remain near the surface without compromising its charge state stability and functional properties is still not well established.  Since the (111) N-terminated diamond slab is the most stable configuration, we perform calculations on the depth dependence for only this system. The 8 nm diamond slab allows to study three different depths: 2~nm, 3~nm, and 4~nm. For each system, we observe the position of the defect states within the gap and the energy proximity to the surface states as shown in Fig. \ref{fig:2}(c). 

Our results show the energy separation between defect levels and the VBM decreases as the NV$^{-}$ center moves closer to the surface (from 1.18 eV at 4 nm to 1.07 eV at 3 nm to 1.0 eV at 2 nm) and similarly, the defect levels and the first unoccupied surface state (from 0.26 eV at 4 nm to 0.08 eV at 3 nm to 0.03 eV at 2 nm), suggesting that surface-induced effects are modifying the defect states more significantly at shallower depths which causes the defect states to be less isolated inside the band gap. As the NV$^{-}$ center becomes shallower (2 nm), the energy associated with these defect states shifts and mixes more strongly with the surface states, leading to partial charge quenching or hybridization. In fact, we see a hybridized surface-bulk state close to the VBM which only appears for the 2 nm depth. The presence of additional surface-induced states (shown as gray blocks) increases as we go deeper into diamond. The splitting between defect levels remains around 1.6 – 1.68 eV for all three depths. Moreover, these findings indicate that NV$^{-}$ centers should ideally be deeper than $\sim$~3 nm to reduce the influence of surface states while maintaining their characteristic energy levels. However, since DFT fails to accurately predict excitation energies, this conclusion must be validated using a more accurate many-body approach.

Since DFT single particle state energies are not to be compared to the physical QP excitations, we will now turn to the MBPT accounting for the long-range (non-local) correlations. These limitations become especially significant in the presence of shallow, localized states interacting with the continuum of bulk or surface states. We compute the QP energies near the Fermi level using the \textit{G$_0$W$_0$} approximation within the stochastic MBPT framework \cite{vlcek2017stochastic, Vlcek2018swift} which efficiently treats large supercells without the prohibitive computational cost of conventional methods \cite{Apelian2024, Romanova2022, Brooks_2020}. Taking our most stable system, the (111) N-terminated diamond slab at a 4 nm depth, we perform the many-body calculations and compare them with DFT to further analyze the positions of the defect states. 

\begin{figure}
    \centering
    \includegraphics[width=4in]{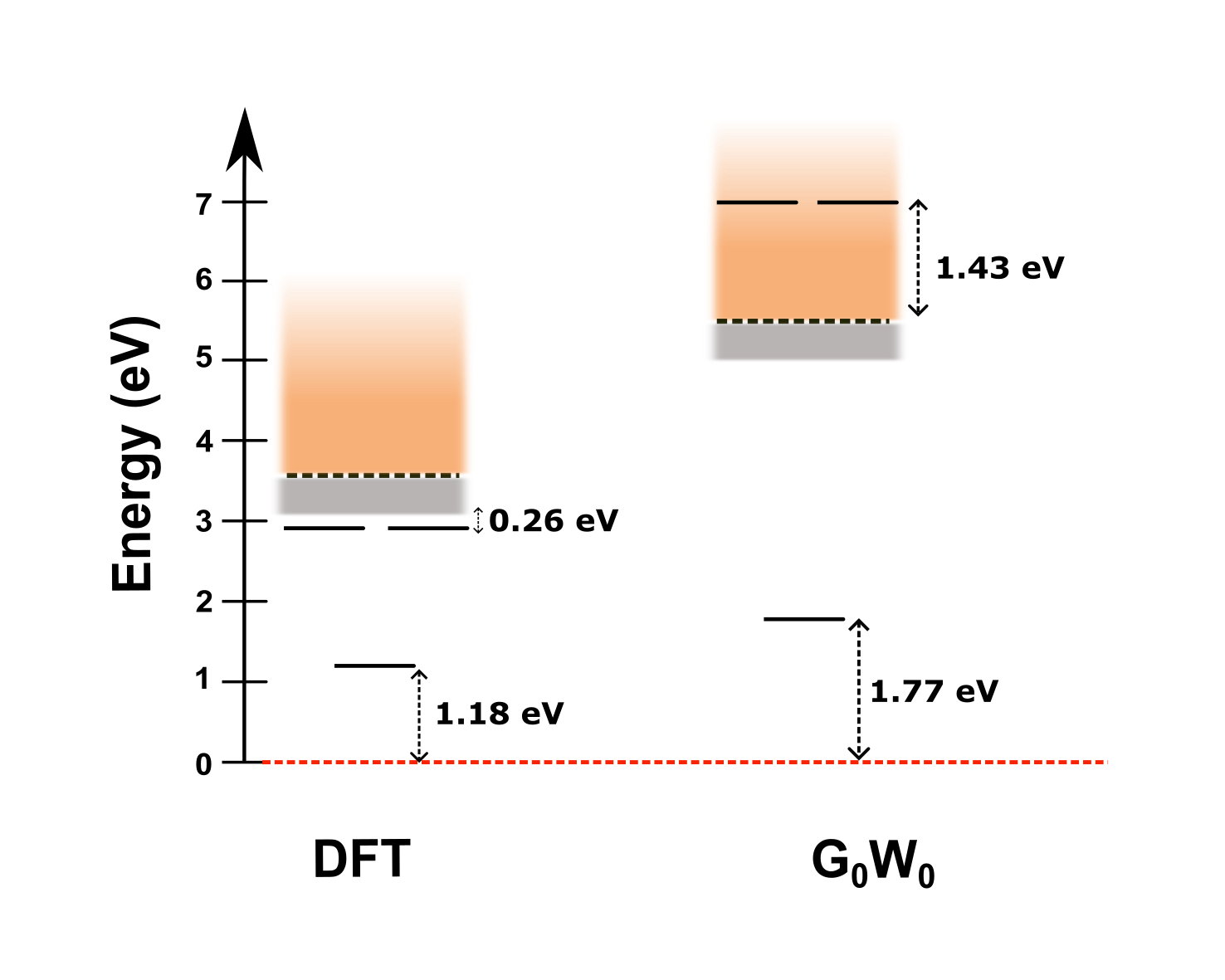}
    \caption{Energy diagrams for the (111) N terminated slab with the defect placed at 4 nm using DFT (left) and \textit{G$_0$W$_0$} (right). The DFT shows a stable configuration whereas the \textit{G$_0$W$_0$} shows the unpaired, degenerate defect states get shifted and reordered to the continuum of unoccupied states (shown as an orange block), significantly interfering with the stable charge configuration. The surface states are shown as gray blocks. } 
    \label{fig:4}

\end{figure}

First, DFT initially predicts three in-gap defect states that suggests charge stability, however, the \textit{G$_0$W$_0$} calculations correct the bands, expanding the band gap and shifting the defect levels. Specifically, the VBM (red dashed lines) and the first defect state gets shifted by 2.28 eV and 1.69 eV, respectively. The band gap opens up to $\sim$~5.56 $\pm$ 0.14 eV which is the result obtained also for the converged bulk limit \cite{Romanova2023} and this value is also close to the experimental value of 5.46 eV \cite{Clark1964} for bulk diamond. This contrasts with the PBE value of 3.51 eV and, therefore, we observe a larger energy separation between the VBM and the first defect state: 1.18 eV with DFT vs 1.77 eV with \textit{G$_0$W$_0$}. Next, we observe a reordering of the states where the two degenerate in-gap states become unoccupied and the defect no longer holds its expected electronic configuration. This reordering is significant where the energy difference between the CBM (shown as black dashed lines) and the defect states are 1.43 eV. This also suggests that charge stability is compromised, likely because the defect is too close to the surface where charge fluctuations and surface interactions play a role. In terms of emission, any photoexcitation will make the additional electron jump to the continuum of empty states (shown as an orange block) rather than to another unoccupied localized state, implying that upon excitation, the NV$^{-}$ center is prone to photoionization, losing its charge state and potentially becoming NV$^{0}$ or another unstable charge configuration. So, while DFT suggests stability, the many-body calculations reveal it to be otherwise, meaning DFT alone is insufficient to predict NV$^{-}$ stability near surfaces. This also suggests a depth of 4 nm is insufficient for NV$^{-}$ charge state stability in an N-terminated (111) slab; experimental efforts to place NV$^{-}$ centers at similar depths might observe fluctuating charge states or reduced optical coherence. \footnote{It is important to note that in this work, we focus on single-particle spectra and do not access the full many-body excitation spectrum, such as the true singlet and triplet transitions. These many-body effects can be captured using downfolding and embedding techniques \cite{Romanova2023, Ma2021}, which allow for explicit modeling of the excited-state manifold. However, in our case, applying such methods to the 4 nm (111) N-terminated slab becomes, in practice, meaningless due to the apparent instability of the system.}

In conclusion, we studied the electronic levels in quantum defects of the NV$^{-}$ centers embedded into diamond slabs, focusing on two key factors: (i) surface reconstruction and termination effects across different surface orientations, e.g, (100) and (111) planes with hydrogen or nitrogen termination and (ii) defect placement at varying depths within large 8 nm slabs. We aimed to uncover the mechanisms that govern defect-surface interactions in shallow NV$^{-}$ centers and to understand how dynamical charge fluctuations near the surface influence defect stability and functionality. We find that surface termination plays a critical role in preserving in-gap defect states: H-terminated slabs are highly unstable due to their NEA and presence of deep-lying surface states close to the band gap, the (100) N-terminated surface induces strong surface hybridization and mixing with defect states, heavily compromising its stability, and the (111) N-terminated surface provides a more favorable configuration, making it the optimal surface type for NV-based quantum sensors. Furthermore, we observe the defect depth dependence of the slabs at 2 nm, 3 nm, and 4 nm, confirming that the latter is the most stable at the DFT level, however, the more appropriate many-body calculations within the $G_0W_0$ approximation, which incorporate the dynamical correlation effects, change this picture dramatically due to the strong coupling to the polarizable surface states. As discussed above, the many-body corrections are critical to properly describe the defect state positioning, and indicate that NV$^{-}$ centers embedded at $\sim$~4 nm or less are prone to surface-induced ionization. Thus, our many-body results establish that NV$^{-}$ centers must be positioned at depths greater than 4 nm within (111) N-terminated diamond slabs to retain their charge and electronic properties. We believe this work sets a theoretical limit on the minimum viable depth for stable, shallow NV$^{-}$ centers in diamond, which is critical for optimizing their use in near-surface quantum sensing applications.

\begin{suppinfo}
Computational details, construction of slabs, stochastic many-body theory.

\end{suppinfo}

\begin{acknowledgement}
The authors thank Ania Jayich, Lillian Hughes, and Shreyas Parthasarathy for their fruitful discussions. The simulations for the large-scale systems, and the analysis were supported by the National Science Foundation (NSF) CAREER award through grant No. DMR-1945098. We acknowledge funding from the UC Office of the President within the Multicampus Research Programs and Initiatives (M23PR5931) supporting some of the calculations performed in this work. Use was made of computational facilities purchased with funds from the NSF (CNS-1725797) and administered by the Center for Scientific Computing (CSC). The CSC is supported by the California NanoSystems Institute and the Materials Research Science and Engineering Center (MRSEC; NSF DMR 2308708) at UC Santa Barbara. A.A. was supported by the NSF Graduate Research Fellowship under Grant No. (2139319), and, in part, by National Science Foundation through Enabling Quantum Leap: Convergent Accelerated Discovery Foundries for Quantum Materials Science, Engineering and Information (Q-AMASE-i) award number
DMR-1906325.

\end{acknowledgement}

\bibliography{bib}

\end{document}